\documentclass[12pt]{article}
\usepackage{geometry}
\usepackage{setspace}			
\usepackage{authblk}			     
\usepackage{natbib}			        
\usepackage[hidelinks]{hyperref}	
\usepackage[labelfont=bf]{caption}
\usepackage{amsmath,amsfonts,amssymb}	
\usepackage{bm}				    
\usepackage{optidef}			

\usepackage{graphicx}			
\usepackage{subcaption}			
\graphicspath{{./}}				

\usepackage{longtable}			
\usepackage{booktabs}			
\usepackage{multirow}			
\usepackage{multicol}			

\usepackage{color}				
\newcommand{\ro}{\text{o}}
\newcommand{\rn}{\text{n}}
\newcommand{\ter}{\text{er}}
\newcommand{\tu}{\text{u}}
\newcommand{\td}{\text{d}}
\usepackage{ulem}				
\usepackage[bb=stix]{mathalpha} 

\usepackage{float}              
\usepackage[section]{placeins}  

\setcitestyle{authoryear,round}

\DeclareMathOperator*{\argmin}{arg\,min} 

\begin{document}

\title{Asset Pricing in the Presence of Market Microstructure Noise}

\author{Peter Yegon}
\author{W. Brent Lindquist\thanks{Corresponding author, brent.lindquist@ttu.edu}}
\author{Svetlozar T. Rachev}
\affil{Department of Mathematics \& Statistics, Texas Tech University, Lubbock, TX, USA}

\date{}	
\maketitle
\begin{abstract}
We present two models for incorporating the total effect of market microstructure noise
into dynamic pricing of assets and European options.
The first model is developed under a Black--Scholes--Merton, continuous--time framework.
The second model is a discrete, binomial tree model developed as an extension of the
static Grossman--Stiglitz model.
Both models are market complete, providing a unique equivalent martingale measure
that establishes a unique map between parameters governing the risk--neutral and real--world
price dynamics.
We provide empirical examples to extract the coefficients in the model, in particular
those coefficients characterizing the influence of the microstructure noise on prices.
In addition to isolating the impact of noise on the volatility, the discrete model enables
us to extract the noise impact on the drift coefficient.
We provide evidence for the primary microstructure noise we believe our empirical
examples capture.
\end{abstract}

\noindent
{\bf Keywords:} market microstructure noise; asset pricing; option pricing; Grossman--Stiglitz model; binomial tree

\doublespacing

\section{Introduction}\label{sec:intro}
Market microstructure effects (market frictions) are collectively viewed as noise affecting 
market--efficient (fundamental) prices.\footnote{
	Microstructure noise introduces further uncertainty into the model, 
	representing factors that are not easily observed or estimated by traders.
	As a result, traders must make decisions based on incomplete information, 
	which affects their ability to price assets accurately \citep{Zhang_2003}.
}
The extent to which, and the time scales on which, these effects impact price is a matter of
continued investigation \citep[for an early survey see][]{Madhavan_2000}.
The problem is aggravated by the fact that the market--efficient price process, which is unobservable,
is unknown.
In this context, the question of the existence of an efficient market also arises
\citep[see][]{Grossman_1980,Vives_2014}.
The classical Black--Scholes--Merton (BSM) price dynamics based on continuous--time geometric Brownian
motion is well--known to be too simplistic to give the appropriate fundamental prices.
Many models, both in continuous and discrete time, have been (are being, and will continue to be)
developed to attempt to capture the stylized facts (volatility clustering, skewness, heavy tails) of
empirical price returns \citep{Cont_2001}.
These stylized facts result largely from macroeconomic factors (market shocks) but undoubtedly have
a component due to microstructure noise \citep[see, e.g.,][]{Lee_2012}.
Needless--to--say, disentangling the components of this collective view is difficult and perhaps somewhat
subjective.

A foundational effort was made by \cite{Roll_1984} in relating the bid--ask spread to the first--order
serial covariance of price changes.
The monograph by \cite{Hasbrouck_2007} describes several discrete--time
empirical market microstructure models which build upon Roll’s bid--ask model.
The models are designed to capture, in various ways, the price formation process, incorporating
the sequence of actions and reactions between market makers and traders.
The impact of microstructure noise on price volatility has been a subject of continued investigation
\citep[see, e.g.,][]{Frey_1997,Bandi_2006,Hansen_2006}.
A particular area of concentration, where noise effects are expected to dominate the volatility,
is high frequency trading \citep{Zhang_2003,Ait-sahalia_2011}.

While the literature on the modeling of market frictions on pricing is too extensive to adequately cover,
we note studies of trader information asymmetry \citep[][Chapters 3--6]{O'Hara_1995},
transaction costs \citep{Leyland_1985,Kabanov_1997}, dynamic hedging \citep{Frey_1997},
and liquidity \citep{Cetin_2004,Ait-sahalia_2009}.

The simplest form of market microstructure noise
is defined as a sequence of independent, identically distributed (iid) random variables
$\epsilon_{\tau_i}$, $i = \{1, 2, \ldots\}$,
defined such that the observed market log--price $Y_{\tau_i}$ at times $\tau_i = i \Delta t$ is
\citep[Equation (2.1), p. 68]{Ait-sahalia_2014}
\begin{equation}\label{eq2.3}
    Y_{\tau_i} = X_{\tau_i} + \epsilon_{\tau_i},
\end{equation} 
where $X_{\tau_i}$ is the efficient (fundamental) log--price at $\tau_i$.
This simple case assumes the random variables $\epsilon_{\tau_i}$ are independent of the $X$ process and
have finite first two moments, with a mean of zero.
In the literature on high--frequency econometrics, \citep[see e.g.,][Section 2.3.2]{Ait-sahalia_2014}
the microstructure noise is often modeled as an ARMA process independent of the underlying
Brownian motion (or, more generally, the semimartingale that determines the dynamics of $S$). 
Consequently, $S_t$, $t \geq 0$, ceases to be a semimartingale, 
thereby raising concerns about the validity of no--arbitrage pricing under the  fundamental theorem
of asset pricing \citep{Delbaen_1994}.
Assuming that the market microstructure source is driven by an additional Brownian motion leads to market incompleteness,\footnote{\
	This is a similar problem to that of local volatility models, 
	where the existence of two sources of risk leads to market incompleteness;
	see the discussion in \cite{Shirvani_2020}
} 
thereby preventing the hedger from perfectly hedging a short position in the option contract. 

Motivated by \eqref{eq2.3}, in Section~\ref{sec:BSM} we extend the BSM framework so that
the price of a risky asset includes a term representing the total effects of microstructure noise
in a manner such that the market remains complete.
Via a replicating portfolio, we develop the partial differential equation describing the dynamics of a
European option having the risky asset as underlying, and present the Feynman--Kac solution.
In the case of constant coefficients, we show that the option price reduces to the familiar BSM
formula under a changed volatility $\sigma + \epsilon$, 
where $\sigma$ is the classical BSM volatility and $\epsilon$ represents additional volatility
due to the microstructure noise.

Section~\ref{sec:rn_eval} provides an alternative derivation of these results,
employing the risk--neutral valuation framework. 
The concept of risk--neutral valuation is central to asset pricing theory,
 providing a framework under which arbitrage opportunities are absent
and prices can be determined based purely on the present value of expected future payoffs.

In Section~\ref{sec:emp_eps}, we present an empirical evaluation of an $\epsilon(T,K)$ surface
implied by prices of (European, cash--settled) call options on the \^{}SPX index.
The empirical surface is computed assuming that the BSM component of the volatility, $\sigma$,
is given either as a simple historical volatility or computed using an ARMA--GARCH model.
In effect, empirically we break the volatility noise term into two pieces,
the noise affecting the spot price of the underlying (which is captured in $\sigma$),
and additional noise generated by trades made by the hedger holding the short position in the option.
Using a simple historical volatility computation for $\sigma$ produces a spot price volatility reflecting
average microstructure noise.
Use of an ARMA--GARCH model (which additionally makes no assumptions regarding Markov nature of
prices), attempts to capture a more accurate description of the microstructure noise component of $\sigma$.

In Section~\ref{sec:DGSM}, we develop an extension of the static model of \cite{Grossman_1980},
which we refer to as the dynamic Grossman--Stiglitz model (DGSM).
Under the DGSM, the drift coefficient and the volatility of the asset's log--return process
are each assumed to consist of the sum of an ``observable'' component and a noise component.
We further assume that the observable and noise components of the drift term are proportional
to each other.
Finally, the DGSM assumes that the return drift is observable at a cost.

The DGSM is developed in continuous time.
However, to avoid the loss of the drift term that occurs when option prices are computed assuming
trading can occur continuously in time,
in Section~\ref{sec:DGSM_bt} we develop a discrete, binomial tree, option pricing version of the DGSM.
By starting in the real world, and transitioning to the risk--neutral world via a replicating,
self--financing portfolio,
the binomial tree model produces a unique equivalent martingale measure which establishes a unique
map between parameters governing the risk--neutral and real--world price dynamics.\footnote{
	Thus, providing a solution to the discontinuity puzzle of option pricing \citep{Kim_2016, Kim_2019}.
}

In Section~\ref{sec:calib}, we describe a method for calibrating the parameters appearing in the discrete
DGSM. 
Empirical estimation of these parameters are illustrated in Section~\ref{sec:p_cal} using the
\^{}SPX data set of Section~\ref{sec:emp_eps}.
A final discussion is presented in Section~\ref{sec:disc}.

\section{BSM Framework Incorporating Market Microstructure Noise}\label{sec:BSM}
We work within a BSM market ($\mathcal{S}$, $\mathcal{B}$, $\mathcal{C}$) consisting of
a risky asset (stock) $\mathcal{S}$,
riskless asset $\mathcal{B}$, and European contingent claim (option) $\mathcal{C}$.
Consider the stochastic basis $(\Omega, \mathbb{F} = \{ \mathcal{F}_t, t \geq 0 \} 
\subset \mathcal{F},\mathbb{P})$ 
on a complete probability space $(\Omega, \mathcal{F}, \mathbb{P})$
generated by a standard Brownian motion $B_t$, $t \geq 0$, on $(\Omega, \mathcal{F}, \mathbb{P})$
with $\mathcal{F}_t = \sigma(B_u, 0 \leq u \leq t)$, $t \geq 0$.
The risky asset $\mathcal{S}$ has price dynamics $S_t$, $t \geq 0$,
determined by the continuous diffusion process,
\begin{equation}\label{eq2.1}
     dS_t = \mu_t S_t dt + \sigma_t S_t dB_t + \epsilon_t S_t dH_t, \quad S_0 > 0,
\end{equation}
where $\mu_t = \mu(S_t, t) \in \mathbb{R}$, $\sigma_t = \sigma(S_t, t) > 0$,
and $\epsilon_t = \epsilon(S_t, t) \in \mathbb{R}$.\footnote{
	The regularity conditions for $\mu_t$, $\sigma_t$ and $\epsilon_t$, $t \geq 0$, 
	are given in  \citet[Section 5G]{Duffie_2001}.
} 
The added term $\epsilon_t S_t dH_t$ reflects the instantaneous market microstructure effects.
The process $H_t$ is
\begin{equation}\label{eq2.2}
    \begin{aligned}
	H_t &= \int_0^t \text{sgn}(B_s) dB_s = |B_t| - L_t,  \quad t \geq 0,\\
	dH_t &= \text{sgn}(B_t) dB_t,
    \end{aligned}
\end{equation}
where $L_t$ is the local time and $\text{sgn}(a)$ is defined as 1, 0, or $-$1 if $a$
is greater than, equal to, or less than zero, respectively.
This representation of $H_t$ is derived from Tanaka's formula\footnote{
        The process $L_t$, {$t \geq 0$},
        represents the local time that the Brownian motion spends at 0 over the interval \([0, t]\), 
        and it is defined as
        $L_t = \lim_{\epsilon \downarrow 0} \frac{1}{2\epsilon} \text{Leb}\{s \in [0, t] \,|\, B_s \in (-\epsilon, \epsilon)\}$,
        where $\text{Leb}$ denotes the Lebesgue measure. 
         The process $H_t = \int_0^t \text{sgn}(B_s) \, dB_s$, $ t \geq 0$,
         has the same distribution as a standard Brownian motion, 
 }
 \citep[see][Chapter 7]{Chung_1990}.
We note that the alternate choice
$H_t = \int_0^t \text{sgn}(B_s^{(\text{Noise})}) dB_s^{(\text{Noise})}$,
where $B_t^{(\text{Noise})}$ is a second Brownian motion possible correlated with $B_t$
(such that  $dB_t^{(\text{Noise})} dB_t = \rho dt$, $\rho \in [0, 1)$),
leads to market incompleteness.
The term $dS_t = \mu_t S_t dt + \sigma_t S_t dB_t$ is viewed as the dynamics of the efficient
(fundamental) asset price.

The observed cumulative return process \citep[][p. 106]{Duffie_2001}
$R_t^{(\text{obs})}$ has the dynamics
\begin{equation}\label{eq2.4}
    \begin{aligned}
	dR_t^{(\text{obs})} &= \frac{dS_t}{S_t} = dR_t^{(\text{eff})} + dR_t^{(\text{MM})}, 
			\quad t \geq 0, \quad R_0^{(\text{obs})} = 0, \\
	dR_t^{(\text{eff})} &= \mu_t dt + \sigma_t dB_t, \quad R_0^{(\text{eff})} = 0,\\
	dR_t^{(\text{MM})} &= \epsilon_t dH_t = \epsilon_t \text{sgn}(B_t) dB_t, \quad R_0^{(\text{MM})} = 0.
    \end{aligned}
\end{equation}
The term, $dR_t^{(\text{eff})}$ defines
the dynamics of the efficient (fundamental) cumulative return process;
the market microstructure (MM) noise is represented by $dR_t^{(\text{MM})}$
Given the numerous sources of noise \citep[see, e.g.,][]{Easley_2003},
we interpret $dR_t^{(\text{MM})}$ as the aggregate effect of these noises.
Note that the terms $\text{sgn}(B_t)$ are random signs, albeit dependent on the uncertainty defined by the
Brownian motion $B_t$ for $t \geq 0$ in $dR_t^{(\text{eff})}$.

The riskless asset $\mathcal{B}$ has the usual dynamics,\footnote{
	The regularity conditions for $r_t$, for $t \geq 0$, are described in \citet[Section 5G]{Duffie_2001}.
} 
\begin{equation}\label{eq2.6}
	d\beta_t = r_t \beta_t dt,     \quad \beta_0 > 0,     \quad r_t = r(S_t, t).
\end{equation}

The option $\mathcal{C}$ has the price $C_t = f(S_t, t)$, where $f(x, t)$, $x  > 0$, $t \in [0, T]$, 
has continuous partial derivatives $\frac{\partial^2 C(x, t)}{\partial x^2}$ and $\frac{\partial C(x, t)}{\partial t}$,
on $t \in [0, T) $ and $g(x)$, $x \in \mathbb{R}$.
Here $T$ is the maturity time $T$ and the option's payoff is $C_T = g(S_T)$ for some continuous function
$g  :  \mathbb{R} \to \mathbb{R}$.
From Itô’s formula,
\begin{equation}\label{eq2.7}
    \begin{aligned}
	dC_t = df(S_t, t) &= \left[ \frac{\partial f(S_t, t)}{\partial t} 
		+ \frac{\partial f(S_t, t)}{\partial x} \mu_t S_t + \frac{1}{2} \frac{\partial^2 f(S_t, t)}{\partial x^2} (\sigma_t + \epsilon_t)^2 S_t^2 \right] dt\\
		& + \frac{\partial f(S_t, t)}{\partial x} [\sigma_t + \epsilon_t \text{sgn}(B_t)] S_t dB_t.
    \end{aligned}
\end{equation}
Note that in the term $ (\sigma_t + \epsilon_t)^2 $ we have absorbed $\textrm{sgn}(B_t)$ into the sign
of $\epsilon_t$.
Thus $\epsilon_t$ need not be a positive quantity in that the noise term (at certain times $t$) can act
to reduce the overall volatility (relative to $\sigma$).
We assume there exists a self--financing strategy $(a_t, b_t)$, $t \geq 0$, 
such that the option price is obtained through a replicating portfolio
\begin{equation}\label{eq:Ct_rep}
	C_t = a_t S_t + b_t \beta_t.
\end{equation}
Following the usual steps for the BSM partial differential equation (PDE) \citep[see, e.g.,][Chapter 5]{Duffie_2001},
we obtain
\begin{equation}\label{eq2.10}
	\frac{\partial f(x,t)}{\partial t} +  r(x,t) x \frac{\partial f(x,t)}{\partial x}
	+ \frac{1}{2}  [\sigma(x,t) + \epsilon(x,t)]^2 x^2 \frac{\partial^2 f(x,t)}{\partial x^2} -r(x,t) f(x,t) = 0,
\end{equation}
subject to the boundary condition
\begin{equation}\label{eq2.11}
	f(x,T) = g(x), \quad x > 0.
\end{equation}

The Feynman--Kac solution to \eqref{eq2.10}, \eqref{eq2.11} is
\begin{equation}\label{eq2.13}
	f(x,t) = \mathbb{E}^Q\left[e^{-\int_t^T r(Z_s,s) \, ds} \, g(Z_T) \, \big| \, Z_t = x \right],
\end{equation}
where $Z$ is an  Itô process satisfying\footnote{
    The regularity conditions for $ r(x,t) $ and $\sigma(x,t) + \epsilon(x,t)$, for $ x > 0 $ and $ t \in [0,T] $
    are described in \citet[Appendix E]{Duffie_2001}.
}
\begin{equation}\label{eq2.12}
	dZ_s = r(Z_s, s) ds + \left[\sigma(Z_s, s) + \epsilon(Z_s, s)\right] dB_s^Q, \quad s \in (t, T], \quad  Z_t = x.
\end{equation}
In \eqref{eq2.12}, $ B_t^Q, t \geq 0 $, denotes a standard Brownian motion that generates a stochastic basis
$(\Omega, \mathcal{F}^Q = \{ \mathcal{F}_t^Q, t \geq 0 \} \subset \mathcal{F}, P^Q) $ on a complete probability space 
$(\Omega, \mathcal{F}^Q, P^Q)$,
with $ \mathcal{F}_t^Q = \sigma(B_u^Q, 0 \leq u \leq t)$, $t \geq 0$.

In the constant coefficient case, $ r(x,t) = r $, $ \sigma(x,t) = \sigma $, and $ \epsilon(x,t) = \epsilon $,
the PDE \eqref{eq2.10} becomes
\begin{equation}\label{eq2.17}
	\frac{\partial f(x,t)}{\partial t} +  r x \frac{\partial f(x,t)}{\partial x}
	+ \frac{1}{2}  [\sigma + \epsilon]^2 x^2 \frac{\partial^2 f(x,t)}{\partial x^2} -r f(x,t) = 0.
\end{equation}
In this case, the price of the option is that of classical BSM option pricing,
with the replacement of the volatility $\sigma$ by the noise--augmented volatility
$\sigma + \epsilon$.
Hence, given a call option payoff of $C_T = \max(0, S_T - K)$,
the call option solution will be
\begin{equation}\label{eq2.19}
	C(S_t,t) = f(S_t,t) = S_t \Phi(u^+) - e^{-r(T-t)} K \Phi\left(u^-\right), \quad 0 \leq t < T,
\end{equation}
where
\begin{equation}\label{eq2.20}
	u^+ = \frac{\ln\left(\frac{S_t}{K} \right) + \left(r + \frac{(\sigma + \epsilon)^2}{2}\right)(T-t)}{(\sigma + \epsilon) \sqrt{T-t}},
	\quad u^- = u^+ - (\sigma + \epsilon)\sqrt{T-t}\,,
\end{equation}
and $\Phi (\cdot)$ is the cumulative standard normal distribution function.
In this case $B_t$ and $B_t^Q$ are related by $dB_t^Q = dB_t + \theta^{(\epsilon)} dt$, where
\begin{equation}\label{eq2.16}
    \theta^{(\epsilon)} = \frac{\mu - r}{\sigma + \epsilon} > 0.
\end{equation}
is the market price of risk in the presence of the microstructure noise.\footnote{
	As in the classical BSM option pricing, \eqref{eq2.16} guarantees that the market model
	is complete and free of arbitrage.
}

The main issue is to determine the coefficients $\mu_t$, $\sigma_t$ and $\epsilon_t$ in \eqref{eq2.4},
thus extracting the dynamics of the total--noise volatility $\epsilon_t$.
Under the assumption of constant coefficients,
we argue that $\mu$ and $\sigma$ can be determined by the behavior of spot prices, 
while $\epsilon$ can be calibrated using the market values of option contracts.
(See Section~\ref{sec:disc} for a refined discussion of this point.)
We illustrate an empirical evaluation in Section~\ref{sec:emp_eps}.

\subsection{Alternate Risk--Neutral Valuation}\label{sec:rn_eval}
We provide an alternate derivation of \eqref{eq2.13}
using the risk--neutral valuation in complete markets without arbitrage opportunities.\footnote{
	We follow the approach in \citet[Section 6H]{Duffie_2001}.
}
We return to a starting point of a market model with MM noise $(\mathcal{S},\mathcal{B},\mathcal{C})$
with the price dynamics \eqref{eq2.1}, \eqref{eq2.6} and $C_t = f(S_t,t)$, $t \in [0,T)$ with $C_T = g(S_T)$,
on the stochastic basis (``natural world'') $(\Omega, \mathcal{F} = \{\mathcal{F}_t, t \geq 0\} \subset \mathcal{F}, P)$. 
To determine $C_t = f(S_t,t)$ using risk--neutral valuation, 
we consider the discounted process $D_t = S_t/\beta_t$, $t \geq 0$.
By Itô's formula
\begin{equation}\label{eq4.1}
	dD_t  = (\mu_t - r_t) D_t dt + \sigma_t D_t dB_t + \epsilon_t D_t dH_t.
\end{equation}
We search for a standard Brownian motion $B_t^\mathbb{Q}$, $t \geq 0$, on the stochastic basis (``risk--neutral world)
$(\Omega, \mathbb{F}, \mathbb{Q})$ where $\mathbb{Q} \sim \mathbb{P}$,
such that on $(\Omega, \mathbb{F}, \mathbb{Q})$, $dB_t^\mathbb{Q} = dB_t + \theta_t dt$,
and $D_t$, $t \geq 0$, is a martingale;
\begin{equation}\label{eq4.2}
	dD_t = \sigma_t^{(D)} dB_t^\mathbb{Q}.
\end{equation}
From \eqref{eq4.1},\footnote{
	Note that $H_t, t \geq 0$ has the properties of a standard Brownian motion.
	See for example Theorem 4.2 (vi), Theorem 2.3, and Example on p. 76 of \cite{Chung_1990}.
}
\begin{equation}\label{eq4.3}
    \begin{aligned}
	dD_t  &= (\mu_t - r_t) D_t dt + (\sigma_t + \epsilon_t) D_t dB_t \\
		&= [\mu_t - r_t - (\sigma_t + \epsilon_t) \theta_t] D_t dt + (\sigma_t + \epsilon_t) D_t dB_t^\mathbb{Q}.
    \end{aligned}
\end{equation}
Choosing the market price of risk to be\footnote{
	Equation \eqref{eq4.3} represents the no--arbitrage condition for the market model with MM noise.
}
\begin{equation}\label{eq4.4}
	\theta_t = \frac{\mu_t - r_t}{\sigma_t + \epsilon_t} > 0, \quad t \geq 0,\ \mathbb{P}-a.s. 
\end{equation}
then \eqref{eq4.2} holds with $\sigma_t^{(D)} = (\sigma_t + \epsilon_t) D_t$.

Since $S_t = D_t \beta_t$, by  Itô's formula the dynamics of $\mathcal{S}$ on $(\Omega, \mathbb{F}, \mathbb{Q})$ is
\begin{equation}\label{eq4.5}
	dS_t = r_t S_t dt + (\sigma_t + \epsilon_t) S_t dB_t^\mathcal{Q}.
\end{equation}
By \eqref{eq4.4} the market model with MM noise $(\mathcal{S}, \mathcal{B}, \mathcal{C})$ 
is arbitrage--free and complete, and $C_t / \beta_t$, $t \geq 0$ is a martingale on $(\Omega, \mathbb{F}, \mathbb{Q})$. 
Therefore, for $t \in [0, T]$, the risk--neutral valuation of the option contract in the market model with MM noise$(\mathcal{S}, \mathcal{B}, \mathcal{C})$ is 
\begin{equation}\label{eq4.6}
    C_t = E_t^\mathbb{Q} \left( \frac{\beta_t}{\beta_T} C_T \right) 
        = E_t^\mathbb{Q} \left( e^{-\int_t^T r_u \, du} g(S_T) \right).
\end{equation}

Equations \eqref{eq4.6} and \eqref{eq2.13} are identical
as \eqref{eq2.12} and \eqref{eq4.5} are Itô processes with the same stochastic dynamics.

\subsection{Empirical Example:  Implied $\epsilon$ Surfaces}\label{sec:emp_eps}
Assuming constant coefficients, we illustrate the computation of implied $\epsilon$ surfaces from an
empirical data set.
Let $C^{(\text{emp})}(S_t,T,K)$ denote an empirical call option chain having maturity dates $T$ and strike prices $K$.
Let $C^{(\text{th})} \left( S_t,T,K;r,\sigma, \epsilon \right)$ denote the theoretical call option price computed from
\eqref{eq2.19} for the same set of maturity dates and strike prices.
We computed an implied $\epsilon$ surface from the minimization
\begin{equation}\label{eq:IE}
\epsilon^{(\text{imp})}(t;T,K) = \argmin_\epsilon
	\left (
		\frac{C^{(\text{th})} \left( S_t,T,K;r,\sigma, \epsilon\right) - C^{(\text{emp})}(S_t,T,K) }
			{C^{(\text{emp})}(S_t,T,K)}
	\right )^2 ,
\end{equation}
Specifically, we illustrated an implied $\epsilon$ surface using a call option chain based on the S\&P 500
index (\^{}SPX)\footnote{
	As options on U.S stocks and ETFs are American--style, finding European--type options
	based on a U.S. instruments is limited to cash--settled options on indexes.
	As American and European call options are priced the same for non--dividend paying stocks,
	we could have used options based on U.S. stocks that have never paid dividends.
	Well--known examples include Amazon (AMZN), Alphabet (GOOGL), Meta Platforms (META),
	Netflix (NFLX) and Berkshire Hathaway (BRK-B).
	To ensure broad market exposure, we chose to use call options on the \^{}SPX index.
}
 for $t = $ 21 April 2025.\footnote{
	Source: Cboe. Accessed 21 April 2025 at 8:01 PM EST.
}
The risk free rate $r$ was provided by the US Treasury daily 10--year par yield curve rate for $t$.\footnote{
	Source: US Treasury. Accessed 21 April 2025 at 8:09 PM EST.
}
We considered two cases: where $\sigma$ in \eqref{eq:IE} is obtained as the historical volatility $\sigma^{(\text{hist})}$
over a historical window $(t-W,t]$,  $W = 1,008$ days, of returns;
and where $\sigma$ was computed by fitting an ARMA--GARCH model to the historical returns.
While $\sigma^{(\text{hist})}$ might be considered a natural estimator for the volatility parameter required in \eqref{eq:IE},
an ARMA--GARCH model should be superior in capturing any ``stylized facts''  \citep{Cont_2001} of the return history,
and therefore produce a better estimator of the volatility at any time $s \in (t-W,t]$.
We utilized an ARMA(3,3)--GARCH(1,1) model,\footnote{
	The ARMA parameters $p = q = 3$ and GARCH parameters $m = n = 1$ were the smallest values
	for which the fitted coefficients were deemed sufficiently significant (see Table~\ref{tab:AGfit}).
}
\begin{equation}\label{eq:AG}
	\begin{aligned}
		r_s &= \phi_0 + \sum{i=1}^3\phi_i r_{s-i} +  a_s + \sum_{j=1}^3\theta_j a_{s-j},\\
		a_s &= \sigma_s \xi_s,\\
		\sigma_s^2 &= a_0 + a_1 a_{s-1}^2 + \beta_1 \sigma_{s-1}^2,
	\end{aligned}
\end{equation}
fit to the window $s \in (t-W,t]$ of return values.
The innovations $\xi_s$ in \eqref{eq:AG} were assumed to be $t-$distributed having degrees--of--freedom
$\nu$.
Table~\ref{tab:AGfit} presents the fitted coefficients and their $p$ values.
Only the constant GARCH parameters $a_0$ was not significant at either the 0.1\% or 1\% level.

\begin{table}[H]
	\centering
	\caption{Parameter values for the ARMA--GARCH fit \eqref{eq:AG} to the historical return series of \^{}SPX.
	$p-$values presented in parenthesis. *** denotes a $p-$value $< 0.001$.}
	\label{tab:AGfit}
	\begin{tabular}{cccc }
	\toprule
	$\phi_0$ & $\phi_1$ & $\phi_2$ & $\phi_3$\\
	$6.09\cdot 10^{-4}$ (0.007) & 0.506 (0.001) & 0.558 (***) & $-$0.728 (***) \\

	\omit & $\theta_1$ & $\theta_2$ & $\theta_3$\strut\\
	\omit & $-$0.506 (0.002) & $-$0.601 (***) & 0.711 (***)\strut\\

	$a_0$ & $a_1$ & $\beta_1$ & $\nu$\\
	$1.77\cdot10^{-6}$ (0.2) & 0.887 (***) & 0.106 (***) & 7.67 (***)\\
	\bottomrule
	\end{tabular}
\end{table}

Specifically, in the second case $\sigma$ in \eqref{eq:IE} was determined from the ARMA--GARCH value
of $\sigma^{(\textrm{AG})} = \sigma_s$ at $s = t$ (that is, for 21 April 2025).
The values obtained were $\sigma^{(\textrm{hist})} = 0.0112$ and $\sigma^{(\textrm{AG})} = 0.0292$.
We refer to the implied $\epsilon$ surfaces computed from the two cases as
$\epsilon^{(\text{imp,hist})}(t;T,K)$ and $\epsilon^{(\text{imp,AG})}(t;T,K)$, respectively.

\begin{figure}[H]
	\centering
	\includegraphics[width=0.49\textwidth]{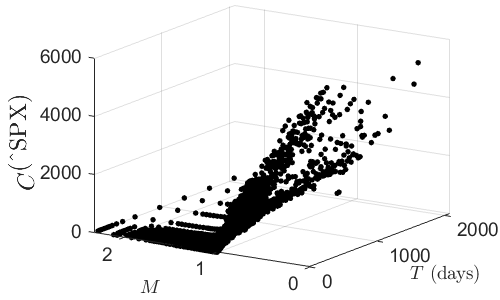}%
	\caption{3D scatter plot of the \^{}SPX call option chain of 21 April 2025.}
	\label{fig:C_emp}
\end{figure}

\begin{figure}[H]
	\centering
	\includegraphics[width=0.49\textwidth]{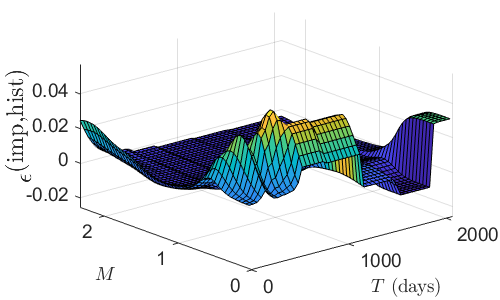}%
	\includegraphics[width=0.49\textwidth]{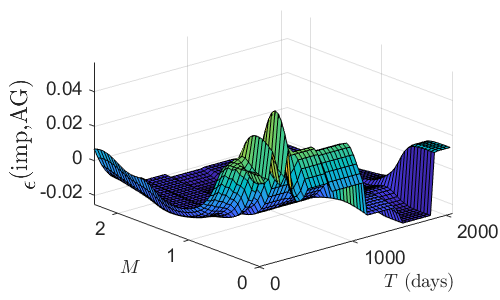}\\
	\caption{Implied $\epsilon$ surfaces.}
	\label{fig:IE}
\end{figure}

Fig.~\ref{fig:C_emp} displays a 3D scatter plot of the prices $C^{(\textrm{\^{}SPX})}$ as a function
of maturity time $T$ and moneyness $M = K/S_0$ for the \^{}SPX call option chain of 21 April 2025.\footnote{
	Option chain data were cleaned by removing data for which
	both the volume and open interest were zero,
	as well as data listed with a zero option price.
}
Fig.~\ref{fig:IE} presents smoothed surface plots\footnote{
	Smoothing performed using a Gaussian kernel on each data point.
	Prior to plotting, the implied $\epsilon$ values were winsorized at the 99\% quantile value.
}
of the resultant implied $\epsilon$ surfaces computed using the two methods of determining $\sigma$.
As $\sigma$ in \eqref{eq:IE} is a constant, values of implied $\epsilon$ ``pick up'' the well--known
volatility smile of the BSM model, as evidenced in Fig.~\ref{fig:IE}.
Reflecting its nature as noise, implied $\epsilon$ values for in--the--money values show significant
variability.

\begin{figure}[H]
	\centering
	\includegraphics[width=\textwidth]{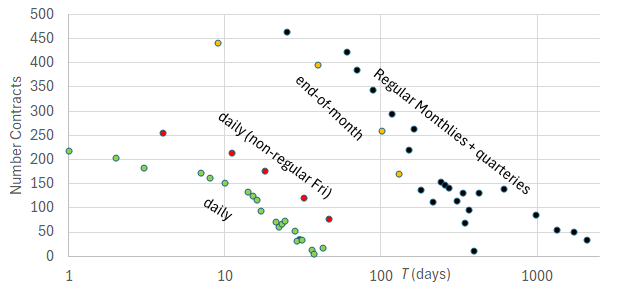}%
	\caption{Number of \^{}SPC call contracts for each maturity time $T$ on 21 April 2025.}
	\label{fig:T_contracts}
\end{figure}

Fig.~\ref{fig:T_contracts} plots the number of \^{}SPX call contracts for each maturity time $T$.
The plot indicates two distinct data subsets; the first (green and red points) consisting of daily contracts,
the remainder (yellow and black points) consisting of regular monthly (closing on the third Friday of each month),
quarterly (closing on the last trading day of each financial quarter) and end--of--month contracts (specific to
options written on indices).
Parenthetically we note the larger number of daily contracts (red points) that mature on a (non--regular) Friday
compared to the number of daily contracts (green points) maturing on a Monday through Thursday,
reflecting traditional close--out of weekly positions, institutional hedging cycles, and the fact that Friday
options have the highest liquidity.

\begin{figure}[H]
	\centering
	\includegraphics[width=0.49\textwidth]{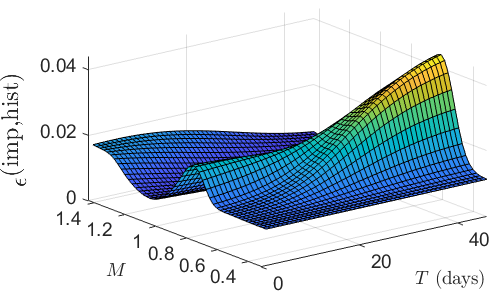}%
	\includegraphics[width=0.49\textwidth]{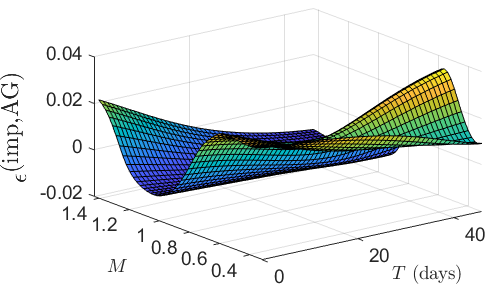}\\
	\includegraphics[width=0.49\textwidth]{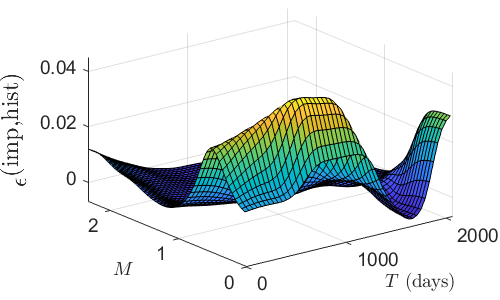}%
	\includegraphics[width=0.49\textwidth]{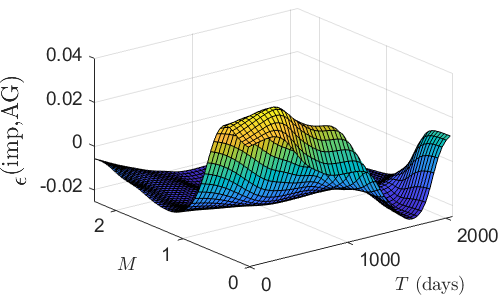}%
	\caption{Implied $\epsilon$ surfaces computed for the four (short/long maturity,
		historical/ARMA--GARCH volatility) combinations.}
	\label{fig:IE_LS_smth}
\end{figure}

Suspecting that the pricing of the shorter term daily contracts is different from the longer term (monthly/quarterly)
contracts, we computed implied $\epsilon$ values separately for these two data subsets.
This led to four combinations: consisting of whether the implied value was computed for the short or long term
maturity set using either the historical of ARMA--GARCH volatility.
Fig.~\ref{fig:IE_LS_smth} plots the implied $\epsilon$ surfaces computed for these four combinations.
The short--term surfaces are much smoother, suggesting that some of the irregularity seen in
Fig.~\ref{fig:IE} was due to  combining two option data sets driven by different pricing.

A few details stand out.
The implied $\epsilon$ ``smile'' is more pronounced when computed with the ARMA--GARCH derived volatility.
For the short--term contracts, there is a pronounced increase in $\epsilon$ as $T$ increases
for values of $M \lesssim 1$.
For the long--term contracts, the pronounced variation in $\epsilon$ occurs deeper in--the--money.

\begin{figure}[H]
	\centering
	\includegraphics[width=0.49\textwidth]{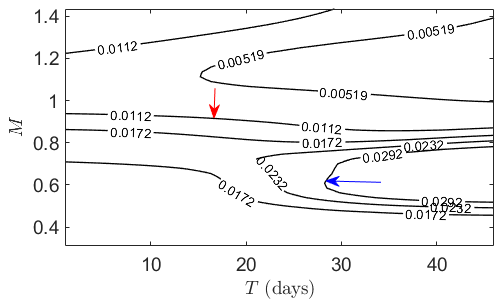}%
	\includegraphics[width=0.49\textwidth]{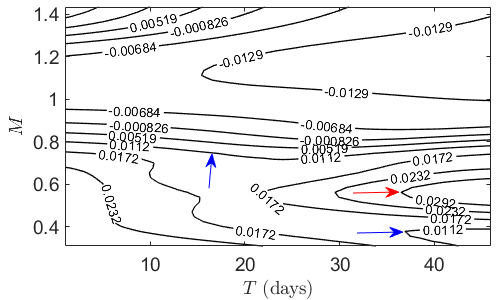}\\
	\includegraphics[width=0.49\textwidth]{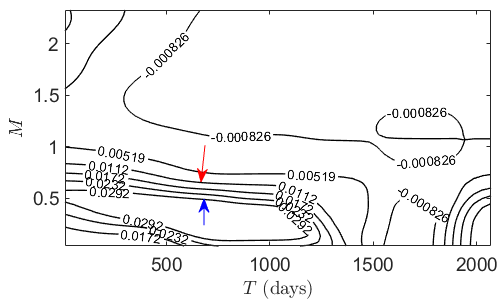}%
	\includegraphics[width=0.49\textwidth]{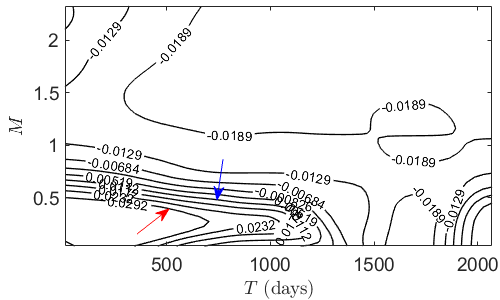}%
	\caption{Contour plots for the $\epsilon$ surfaces of Fig.~\ref{fig:IE_LS_smth}
		as projected on the $T,M$ plane.
		The plots occur in the same order as those in Fig.~\ref{fig:IE_LS_smth}.
	}
	\label{fig:IE_LS_cont}
\end{figure}

Projected contours from the plots in Fig.~\ref{fig:IE_LS_smth} are shown in Fig.~\ref{fig:IE_LS_cont}.
Contours corresponding to the appropriate value of volatility (historical or ARMA--GARCH) from which
the implied $\epsilon$ surface was computed are indicated by red arrow.
Contours corresponding to the other volatility are indicated by a blue arrow.
As already noted from the surface plots, there is a distinctive difference between the contour plots
for the short--term and long--term call option subsets of the data.
For each subset, the contour plots of implied $\epsilon$ derived from historical or ARMA--GARCH volatility
are qualitatively similar, but with a shift in the position of the contours.
When the ARMA--GARCH volatility is used, values of implied $\epsilon$ become (more) negative
in the out--of--the money region.
For the long--term call option data set, values of implied $\epsilon$ are negative in the large--$T$,
in--the--money region as well, regardless of the volatility used in the computation.

\section{The Dynamic Grossman--Stiglitz Model}\label{sec:DGSM}
\cite{Grossman_1980} \citep[see also][]{Vives_2014} extended the rational expectations ideas of Lucas and Sargent
to the case in which the risky asset has a random return
\begin{equation}\label{eq5.1}
	r = r^{(\ro)} + r^{(\rn)},
\end{equation}
consisting of a component $r^{(\ro)}$ that is a return observable at a cost $c > 0$, 
and a component $r^{(\rn)}$ that  is an unobservable noise variable  having mean value
$\mathbb{E}[r^{(\rn)}] = 0$.
As noted by \citet[footnote 1]{Grossman_1980}, $r^{(\ro)}$ can be viewed as a
measurement of $r$ with error.
Thus $r$ in \eqref{eq5.1} is viewed as the return of the market--efficient (fundamental) price
process.
The random pair $(r^{(\ro)}, r^{(\rn)})$ is assumed to be bivariate normally distributed.
We consider a dynamic analogue of the Grossman--Stiglitz model \eqref{eq5.1} -- the DGSM.
To ensure that the DGSM is complete, we assume that $(r^{(\ro)}, r^{(\rn)})$ are determined by a
common source of uncertainty, 
so that the correlation between $r^{(\ro)}$ and $r^{(\rn)}$ is $\pm 1$ .

In the DGSM,  \eqref{eq5.1} is applied to the cumulative return process of the risky asset.
As in \citeauthor{Grossman_1980}, the assumption that traders observe this return at a cost
implies that the observable cumulative return process $R_t^{(\ro)}$, $ t \geq 0 $, follows the dynamics
\begin{equation}\label{eq5.4}
	dR_t^{(\ro)} = \frac{dS_t^{(\ro)}}{S_t^{(\ro)}} = (\mu^{(\ro)} - c) dt + \sigma^{(\ro)} dB_t, \qquad S_0^{(\ro)} > 0,
\end{equation}
where $S_t^{(\ro)}$ is the observable price process of the risky asset and $c > 0$ is the instantaneous cost.
We assume that $\mu^{(\ro)}$ and $\sigma^{(\ro)}$ can be estimated from historical spot trading data.
As the cost $c$ would be revealed in trading of the replicating portfolio used by the hedger taking the
short position in the option contract, $c$ should be calibrated from option data.
In the present paper, we assume $c = 0$ and address the $c \ne 0$ case in the Discussion.
 
We assume the noise cumulative return process $ R_t^{(\rn)}$ , $t \geq 0$, is also determined by
arithmetic Brownian motion,\footnote{
	Following \cite{Grossman_1980}, we assume that $E_t[dR_t^{(\ro)}] = 0$,
	where $E_t[\cdot]$ denotes the conditional expectation at time $t$.
}
\begin{equation}\label{eq5.5}
    dR_t^{(\rn)} = \frac{dS_t^{(\rn)}}{S_t^{(\rn)}} = \mu^{(\rn)} dt + \sigma^{(\rn)} dB_t, \qquad  S_0^{(\rn)} > 0,
\end{equation}
where $S_t^{(\rn)}$, $ t \geq 0$, is the (unobservable) noise price process of the risky asset.

The market--efficient price dynamics of the risky asset are therefore determined by the
total cumulative price process $R_t$ having the dynamics
\begin{equation}\label{eq5.6}
    \begin{aligned}
	dR_t = \frac{dS_t}{S_t} = dR_t^{(\ro)} + dR_t^{(\rn)} 
		& = (\mu^{(\ro)} + \mu^{(\rn)}) dt + (\sigma^{(\ro)} + \sigma^{(\rn)}) dB_t\\
		& = \mu dt + \sigma dB_t, \quad t \geq 0,
    \end{aligned}
\end{equation}
where $\mu = \mu^{(\ro)} + \mu^{(\rn)}$ and $\sigma = \sigma^{(\ro)} + \sigma^{(\rn)}$
are, respectively, the drift coefficient and volatility of the fundamental return process,
and $S_t$, with $S_0 = S_0^{(\ro)} + S_0^{(\rn)}$, is the price process of the risky asset.
We assume $\mu$ and $ \mu^{(\ro)}$ are proportionately related,
\begin{equation}\label{eq:wter}
	\mu = \mu^{(\ro)} w^{(\ter)},
\end{equation}
and $w^{(\ter)} \neq 0$ is unobservable by spot traders but can be calibrated from
option market data.\footnote{
	As we are going to apply a version of the binomial model of \cite{Kim_2016}, 
	the instantaneous drift coefficient is preserved in the option price.
}
Similarly, the noise volatility $\sigma^{(\rn)} $ is unobservable by spot traders,
but implied values can be calibrated from option data.
Using \eqref{eq:wter}, \eqref{eq5.6} can be rewritten
\begin{equation}\label{eq5.7}
	dR_t  = \mu^{(\ro)} w^{(\ter)} dt + (\sigma^{(\ro)} + \sigma^{(\rn)}) dB_t, \qquad t \geq 0,
\end{equation}

Under the DGSM, the riskless asset has price dynamics
\begin{equation}\label{eq5.2}
    d\beta_t = r \beta_t dt,    \quad \beta_0 > 0,     \quad t \geq 0,
\end{equation}
where $r$ is the instantaneous return of the riskless asset.
(As the price $\beta_t$ is riskless, $r$ has no noise component.)

\subsection{The DGSM Binomial Pricing Tree}\label{sec:DGSM_bt}
Under option pricing in continuous time,
the drift $\mu$  disappears, producing an effect known as the discontinuity puzzle in option pricing
\citep{Kim_2016,Kim_2019}.
This makes calibration of $w^{(\ter)}$ impossible.
This puzzle is resolved by assuming that trading instances occur discretely, 
such as when the price dynamics of the riskless asset is based on a binomial pricing tree
\citep[see,][]{Hu_2020a,Hu_2020b, Lindquist_2024}.
Classical binomial pricing models \citep[see, e.g.,][]{Cox_1979, Jarrow_1983, Hull_2012}
embed the discontinuity puzzle by assuming that the option price is independent of the instantaneous
mean return $\mu$. 
Using the approach of \cite{Kim_2016}, we develop a binomial version of DGSM option pricing
which preserves the drift parameter in \eqref{eq5.7}.

For every $n \in \mathbb{N}$, let $\xi_{(k,n)}$, $k = 1, 2, \ldots, n$,
represent iid Bernoulli random variables with
$\mathbb{P}(\xi_{(k,n)} = 1) = 1 - \mathbb{P}(\xi_{(k,n)} = 0) ={p_n}$
determining the filtration
$$
	\mathbb{F}^{(\rn)} = \left\{ \mathcal{F}_k^{(\rn)} = \sigma(\xi_{(j,n)}, j = 1, \ldots, k),
	\ k = 1, \ldots, n, \ \mathcal{F}_0^{(\rn)} = \{ \emptyset, \Omega \}, \ \xi_{(0,n)} = 0 \right\}
$$
and the stochastic basis $(\Omega, \mathbb{F}^{(\rn)}, \mathbb{P}) $ on the complete probability space $(\Omega, \mathcal{F}, \mathbb{P}) $.
The discrete price of $\mathcal{S}$ is $S_{k\Delta,n}$ at time
$k\Delta$, $k = 0, 1, \ldots, n$, $n \in \mathbb{N} = \{1, 2, \ldots\}$,
where $ \Delta = T/n$, $T$ being a fixed terminal time.
The dynamics of $S_{k\Delta,n}$ is given by
\begin{equation}\label{eq5.8} 
    \begin{aligned}
	S_{(k+1)\Delta,n} &= 
	\begin{cases}
		S_{(k+1)\Delta,n}^{ (\tu)} = S_{k\Delta,n}(1 + u_\Delta), & \text{w.p. } p_n, \\
		S_{(k+1)\Delta,n}^{ (\td)} = S_{k\Delta,n}(1 + d_\Delta), & \text{w.p. } 1-p_n,
	\end{cases}\\
	S_{0,n} &= S_0 > 0.
    \end{aligned}
\end{equation}

The arithmetic return
$r_{(k+1)\Delta,n} = (S_{(k+1)\Delta,n} - S_{k\Delta,n})/S_{k\Delta,n}$,
of  the risky asset satisfies
\begin{equation}\label{eq5.11}
    \begin{aligned}
	r_{(k+1)\Delta,n} &= 
	\begin{cases}
		r_{(k+1)\Delta,n}^{ (\tu)} = u_\Delta, & \text{w.p. } p_n,\\
		r_{(k+1)\Delta,n}^{ (\td)} = d_\Delta, & \text{w.p. } 1 - p_n,
	\end{cases}\\
	r_{0,n} &= 0.
    \end{aligned}
\end{equation}
Following the exposition in \cite{Hu_2020a},
the parameters $u_\Delta$ and $d_\Delta$ are determined by requiring
\begin{equation}\label{eq5.12}
	E[r_{(k+1)\Delta,n}] = \mu \Delta, \quad \text{Var}[r_{(k+1)\Delta,n}] = \sigma^2 \Delta,
\end{equation}
producing
\begin{equation}\label{eq5.13}
	u_\Delta = \mu \Delta + \sqrt{\frac{1 - p_n}{p_n}} \sigma \sqrt{\Delta}, \qquad
	d_\Delta = \mu \Delta - \sqrt{\frac{p_n}{1 - p_n}} \sigma \sqrt{\Delta}.
\end{equation}

The discrete price $\beta_{k,n}$ of $\mathcal{B}$ has the dynamics
\begin{equation}\label{eq5.9}
    \begin{aligned}
 	\beta_{(k+1)\Delta,n} &= \beta_{k\Delta,n}(1 + r\Delta), \quad k = 0, 1, \ldots, n - 1,\\
 	\beta_{0,n} &= 1.
    \end{aligned}
\end{equation}
where $r \geq 0$ is the instantaneous riskless rate.
We require
\begin{equation}\label{eq:no_arb}
	 d_\Delta < r \Delta < u_\Delta
\end{equation}
to ensure no arbitrage. 

Using a standard, self--financing, replicating portfolio argument,
the discrete price  $C_{k\Delta,n}$ of the option $\mathcal{C}$, 
having maturity payoff $C_T = g(S_T)$, is determined by the risk--neutral pricing tree
\begin{equation}\label{eq5.14}
	C_{k\Delta,n} = \frac{1}{1 + r\Delta} \left( q_n C_{(k+1)\Delta,n}^{ (\tu)} + (1 - q_n) C_{(k+1)\Delta,n}^{ (\td)} \right),
	\qquad k = 0, ..., n-1,
\end{equation}
where 
\begin{equation}\label{eq5.16}
	q_n = p_n - \theta \sqrt{p_n(1 - p_n)\Delta},
\end{equation}
with $ \theta = (\mu - r)/\sigma$ being the market price of risk. 
The no--arbitrage condition \eqref{eq:no_arb} guarantees $q_n \in (0,1)$ for all $n \in \mathbb{N}$. 
From \eqref{eq5.14} and \eqref{eq5.16}, the risk--neutral price of the call option is dependent on $\mu$,
which will enable its estimation from option prices.\footnote{
	We note in that classical binomial option pricing models \citep[see,][]{Cox_1979, Jarrow_1983, Hull_2012}
	the risk--neutral option price is independent of both $\mu$ and $p_n$.
	Regardless of the size of $\mu$ or how close $p_n$ is to zero or unity,
	the call option price is the same.
	Thus our formulation \eqref{eq5.14}, \eqref{eq5.16} solves both aspects of this option pricing discontinuity puzzle.
}

As $ n \to \infty $ and $ \Delta = T/n \to 0$, 
the pricing tree represented by \eqref{eq5.8} generates a discrete price process
with values in the Skorokhod space $\mathcal{D}[0,T]$,
which converges weakly to a geometric Brownian motion,
\begin{equation}\label{eq5.17}
	S_t = S_0 e^{( \mu - \sigma^2/2 )t + \sigma B_t}, \quad t \in [0,T],
\end{equation}
where $B_t$,$t \in [0,T]$, is the Brownian motion on $(\Omega, \mathbb{F}, \mathbb{P})$. 
In the risk--neutral world, replacing $p_n$ with $q_n$ in \eqref{eq5.8}
causes the risk--neutral price process to converge weakly in $D[0,T]$ to 
\begin{equation}\label{eq5.18}
	S_t = S_0 e^{( r - \sigma^2/2 )t + \sigma B_t^\mathbb{Q}}, \quad t \in [0,T],
\end{equation}
where $B_t^\mathbb{Q}$, $t \in [0,T]$, is the Brownian motion on $(\Omega, \mathbb{F}, \mathbb{Q})$. 
Thus, in continuous time all information about $p_n$ and $ \mu$ is lost due to the assumption that
the hedger having the short position in the option can trade continuously over time.

\section{Parameter Calibration}\label{sec:calib}
From \eqref{eq5.14}, \eqref{eq5.16}, the price of a call option with maturity $T$ and strike $K$ can be written as
$C^{(\text{bt})}\left(S_0,T,K;r, p_n,\mu,\sigma\right)$, where $\mu$ and $\sigma$ are given by \eqref{eq5.7}.
The rate $r$ is determined by the appropriate riskless asset.
Let $r_{k\Delta}^{(\mathcal{S})}$, $k = -M+1, ..., 0$, denote historical daily returns of the risky asset.
We can estimate the instantaneous mean $\mu^{(\text{o,smpl})}$ and volatility $\sigma^{(\text{o,smpl})}$
from the historical data in the usual manner
\begin{equation}\label{eq6.2}
	\mu^{(\text{o})}  = \frac{1}{M\,\Delta} \sum_{k=1}^{M} r_{k\Delta}^{(\mathcal{S})} , \qquad
	\left(\sigma^{(\text{o})}\right)^2  = \frac{1}{(M-1)\Delta}
			\sum_{k=1}^{M} \left( r_{k\Delta}^{(\mathcal{S})} - \mu^{(\text{o})} \right)^2.
\end{equation}
The probability $p_n$ can also be determined from the historical return series as the observed
fraction of positive returns.

The total instantaneous mean $\mu$ and volatility $\sigma$ can be estimated using market option prices.
Let $C^{(\text{emp})} (S_0,T,K)$ denote a published call option price for $\mathcal{C}$.
Implied values of the pair ($\mu, \sigma$) can be obtained through the minimization
\begin{equation}\label{eq:ms_imp}
	\left( \mu^{\text{imp}}(T,K), \sigma^{\text{imp}}(T,K) \right) = \argmin_{\mu,\sigma}
		\left[ \frac{ C^{(\text{bt})}\left(S_0,T,K;r, p_n,\mu,\sigma\right) - C^{(\text{emp})} (S_0,T,K) } { C^{(\text{emp})} (S_0,T,K) } \right].
\end{equation}
subject to the constraints
\begin{equation}\label{eq:ms_const}
	0 < q_n < 1, \qquad \sigma > 0\,.
\end{equation}
The noise parameters $w^{(\ter)}$ and $\sigma^{(\rn)}$ were computed from the minimization
of the mean absolute errors:
\begin{equation}\label{eq:ws}
    \begin{aligned}
	w^{(\ter)} &= \argmin_{w}
		\text{E}_{T,K} \left[ \,\left| \mu^{(\ro)} w - \mu^{(\text{imp})}(T,K) \right| \,\right],\\
	\sigma^{(\rn)} &= \argmin_{\sigma}
		\text{E}_{T,K} \left[ \,\left|  \sigma^{(\ro)}  + \sigma - \sigma^{(\text{imp})}(T,K) \right| \,\right],
    \end{aligned}
\end{equation}
where $\text{E}_{T,K} [\cdot]$ denotes the expectation over all $K,T$ values.
From the calibrated values $ \mu^{(\ro)}$, $w^{(\ter)}$, $\sigma^{(\ro)}$, and $\sigma^{(\rn)}$ we can
then compute
\begin{equation}\label{eq:final_pars}
		\mu = \mu^{(\ro)} w^{(\ter)}, 
		\qquad  \sigma = \sigma^{(\ro)} + \sigma^{(\rn)}, \\
		\qquad \mu^{(\rn)} = \mu^{(\ro)} \left(w^{(\ter)} - 1\right).
\end{equation}

\subsection{Empirical Calibration using \^{}SPX Data}\label{sec:p_cal}
\begin{figure}[p]
	\centering
	\includegraphics[width=0.49\textwidth]{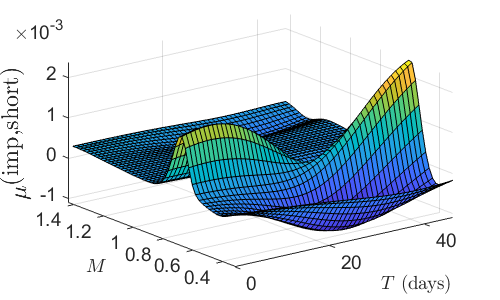}%
	\includegraphics[width=0.49\textwidth]{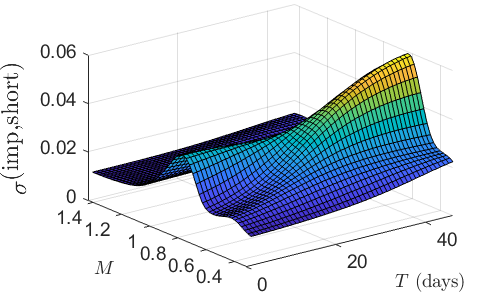}\\
	\includegraphics[width=0.49\textwidth]{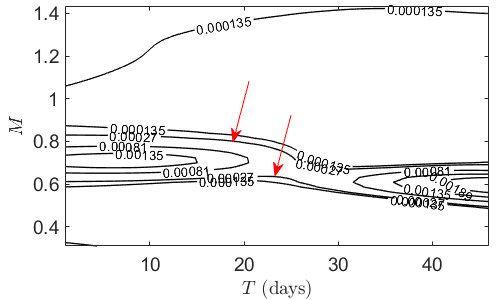}%
	\includegraphics[width=0.49\textwidth]{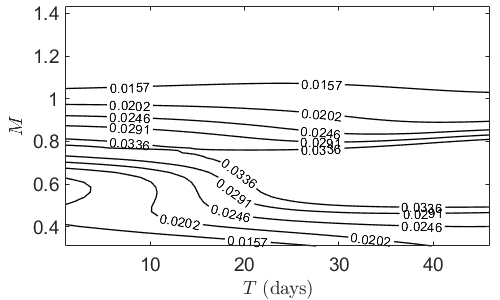}\\
		\includegraphics[width=0.49\textwidth]{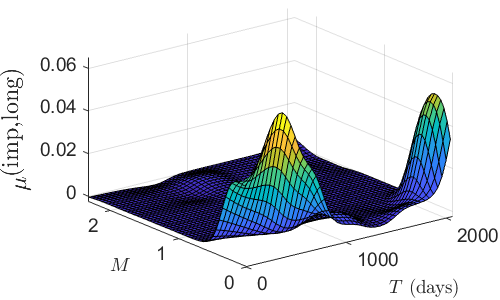}%
	\includegraphics[width=0.49\textwidth]{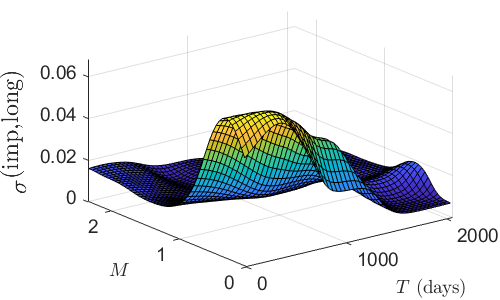}\\
	\includegraphics[width=0.49\textwidth]{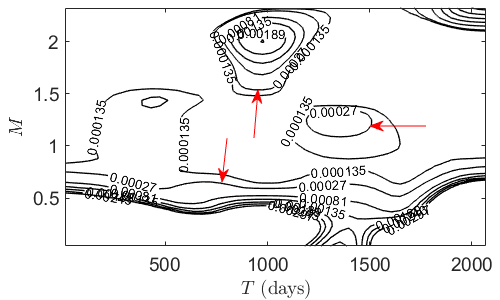}%
	\includegraphics[width=0.49\textwidth]{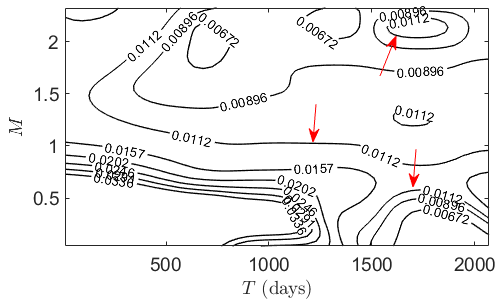}\\
	\caption{Implied $\mu$ and $\sigma$ surfaces and their respective projected
		contour plots for the short-- and long--term call options.}
	\label{fig:ImpMS}
\end{figure}

We illustrate the calibration of these parameters using the same \^{}SPX date set as in Section~\ref{sec:emp_eps}.
Fig.~\ref{fig:ImpMS} shows the implied $\mu$ and $\sigma$ surfaces computed via \eqref{eq:ms_imp}
for the short-- and long--term call options.
Also shown are respective plots of the surface contours projected in the $T,M$ plane.
To quantify the results further, Table~\ref{tab:IQR} summarizes quantile data for each
of the implied surfaces.
The historical value of $\mu$ is larger than 77\% of the implied $\mu$ values computed from
the short--term data, and larger than 88\% computed from the long--term data.
The historical $\sigma$ is smaller than any implied value computed from the short--term data,
while the  historical $\sigma$ falls at the 32'nd percentile of the long--term data implied values.
Thus the call option data (for 21 April 2025) tends to predict a smaller value for the return drift
component than that obtained from the four--year historical window.
While the implied $\mu$ is smaller in the out--of--the money region, there is a difference
between where this occurs in the in--the--money region for the short-- and long--term options.
For this date, the option data tends to predict values of implied $\sigma$ that exceed
the historical data.
For the short--term options, this ``over--prediction'' occurs everywhere in the $T,M$ space.
For the long--term options, this occurs over roughly 68\% of the $T,M$ space.

\begin{table}[tb]
	\centering
	\caption{Quantile values for the implied surfaces presented in Fig.~\ref{fig:ImpMS}}.
	\label{tab:IQR}
	\begin{tabular}{c ccccc cc}
	\toprule
{\bf \strut} & {\bf min} & { $\bm{P_{25}}$} & { $\bm{P_{50}}$} & { $\bm{P_{75}}$} & {\bf max} & {\bf historical} & {\bf historical}\\
	{\strut} & \omit & \omit & \omit & \omit & \omit & {\bf value} & {\bf value}\\
	{\bm \strut} & { $\bm{\times 10^{-2}}$} & {$\bm{\times 10^{-2}}$} & {$\bm{\times 10^{-2}}$} & {$\bm{\times 10^{-2}}$} & { $\bm{\times 10^{-2}}$} &  { $\bm{\times 10^{-2}}$}  & {\bf percentile}\\
	\midrule
	$\mu^{(\text{imp,short})}$ & $-$0.111 & 0.0115 & 0.0115 & 0.0119 & 0.238 & 0.0270 & 77\\
	$\mu^{(\text{imp,long})}$ & $-$0.240 & 0.0115 & 0.0115 & 0.0115 & 0.651 & 0.0270 & 88\\
	$\sigma^{(\text{imp,short})}$ & 1.12 & 1.13 & 1.40 & 1.69 & 5.92 & 1.12 & 0\\
	$\sigma^{(\text{imp,long})}$ & 0.539 & 1.06 & 1.20 & 1.49 & 6.81 & 1.12 & 32\\
	\bottomrule
	\end{tabular}
\end{table}

Table~\ref{table:calibrated_parameters} presents the calibrated parameters computed from the
historical data and from the short-- and long--term call option data.
With the exception of $\sigma^{(\rn)}$, the parameters computed from the short-- and long--term
options show remarkable similarity.
The value of $\mu^{(\rn)}$ is 57\% that of $\mu^{(\ro)}$, but of opposite sign.
The short--term value of $\sigma^{(\rn)}$ is 25\% that of $\sigma^{(\ro)}$,
while the long--term value is 7.5\% of the historical volatility.

\begin{table}[H] 
    \centering
    \caption{Calibrated Parameters}
    \label{table:calibrated_parameters}
    \begin{tabular}{l ccc}
	\toprule
	\strut &  $\bm{\mu^{(\ro)}}$ & $\bm{\sigma^{(\ro)}}$ & $\bm{p_n}$ \\
	historical & $2.70 \cdot 10^{-4}$ & 0.0112 & 0.524\\
	\midrule
	\strut & $\bm{\mu^{(\rn)}}$ & $\bm \sigma^{(\rn)}$ & $\bm{w}^{\bm{(\ter)}}$ \\
	short--term & $-1.55 \cdot 10^{-4}$  & 0.0028   & 0.426\\
	long--term & $-1.55 \cdot 10^{-4}$  & $8.42 \cdot 10^{-4}$  & 0.426  \\
	\bottomrule
    \end{tabular}
\end{table}

\begin{figure}[H]
	\centering
	\includegraphics[width=0.49\textwidth]{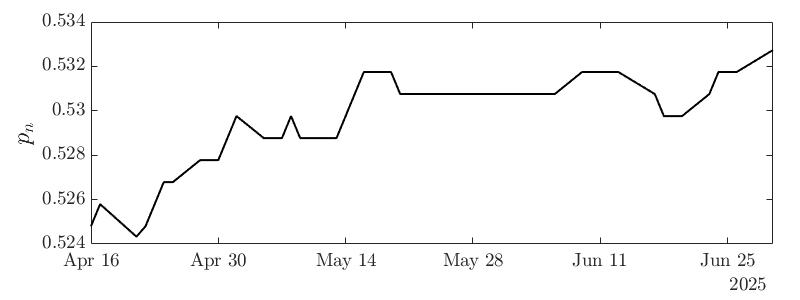}%
	\includegraphics[width=0.49\textwidth]{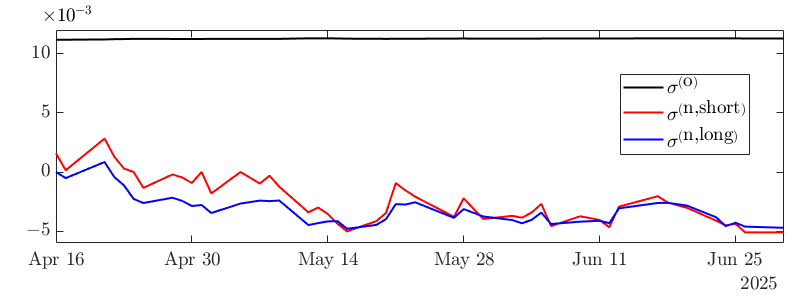}\\
	\includegraphics[width=0.49\textwidth]{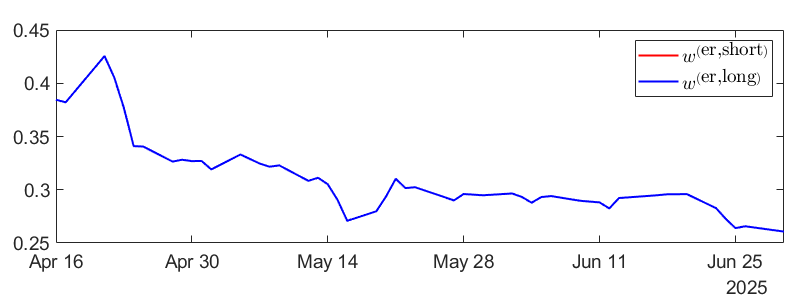}%
	\includegraphics[width=0.49\textwidth]{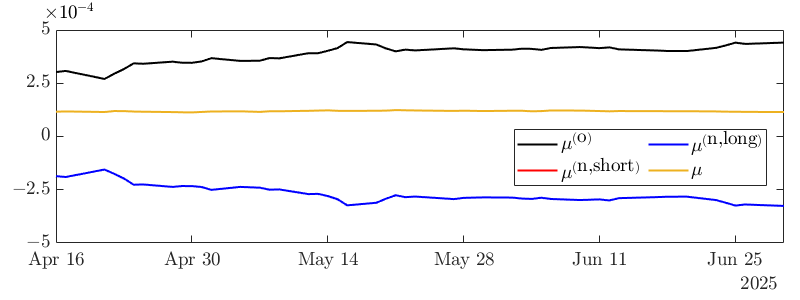}\\
	\caption{Parameter values computed from \^{}SPX option chains covering
		the period 16 April 2025 through 30 June 2025.}
	\label{fig:pars}
\end{figure}

To get a more comprehensive view of the change of these parameters with time,
we redid the computations for these parameters using the data for 46 call option chains
covering the period 16 April 2025 through 30 June 2025.
In each case the historical values were computed using an appropriate window of 1,008
return values.
Fig.~\ref{fig:pars} summarizes the results.
Over this period of time, the historical daily probability $p_n$ for an increase of the \^{}SPX index
increased by 1.5\%.
Except for a brief period after 16 April 2025, both the short-- and long--term vales of $\sigma^{(\rn)}$
were negative.
Consequently, for most of the period, the observable volatility $\sigma^{(\ro)}$ was larger than
 the (unobservable) market--efficient volatility $\sigma$ by the noise--induced amount $\sigma^{(\rn)}$.
 For the few days after 16 April 2025, it appeared that the microstructure noise reduced the observable
 volatility compared to the market--efficient volatility.
 In contrast to the volatility, the short-- and long--term values of $w^{(\ter)}$ were identical;
$w^{(\ter)}$ decreased by a factor of 1/3 over the study period.
Thus the short-- and long--term values of $\mu^{(\rn)}$ were identical.
More importantly $w^{(\ter)} < 1$, indicating that the market--efficient coefficient
$\mu = \mu^{(\ro)} + \mu^{(\rn)} = \mu^{(\ro)} w^{(\ter)}$
was found to have smaller magnitude than either the observable or noise coefficients.

\section{Discussion}\label{sec:disc}
There are many sources of microstructural noise in a market including:
bid--ask bounce; discrete price changes; asynchronous trading; order processing costs;
inventory and liquidity constraints; information asymmetry; latency and stale quotes;
data errors; and algorithmic trading.
There is no averaging (e.g. average price per day, etc.) employed in our empirical data;
rather the data set consists of a regularly spaced (once per trading day) sample of
close--of--market tick prices.
Therefore the microstructure noise inherent in the tick data is not averaged out, rather our
sample captures that microstructural noise present in a daily--spaced sample.

\begin{figure}[H]
	\centering
	\includegraphics[width=0.49\textwidth]{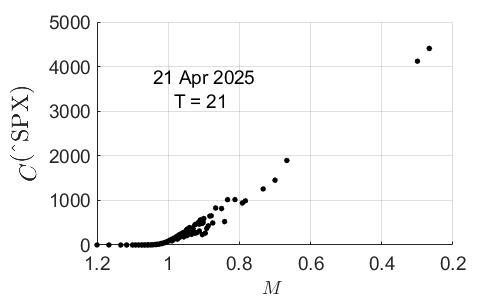}%
	\includegraphics[width=0.49\textwidth]{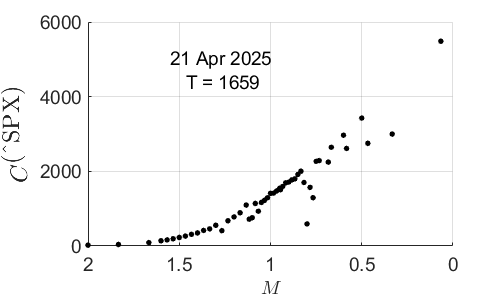}\\
	\caption{Empirical call option price as a function of moneyness for two maturity dates
		for the \^{}SPX call option chain of 21 April 2025.
	}
	\label{fig:CvsK}
\end{figure}

Fig.~\ref{fig:CvsK} presents evidence for the most probable source of such noise in our data set.
The figure plots the empirical call option price as a function of moneyness, $M = K/S_t$ for two maturity
dates, one short--term ($T = 21$ days) and one long--term ($T=1659$ days) for the \^{}SPX call option
chain of 21 April 2025.
The results are representative for our entire data set.
The non--monotonicity of the option prices as $M$ decreases (moves into--the--money) is largely
due to stale ``Last Price'' values resulting from asynchronous trading.\footnote{
	The data had been cleaned to remove obvious data entry errors such as zero call prices.
	Additionally, all contract entries where the volume and open interest were both zero were excluded.
}
We postulate that stale option prices due to asynchronous trading was the predominant source
of microstructure noise in the data set.

We address the issue of including a non--zero cost $c$ in \eqref{eq5.4}.
As this cost would be revealed in trading of the replicating portfolio used by the hedger taking
the short position in the option contract, it should be a function of time to maturity and strike price.
To estimate this cost, the first equation in \eqref{eq:ws} should be modified to
\begin{equation}\label{eq:wc}
	\left\{w^{(\ter)}, \alpha_c\right\} = \argmin_{w,\alpha}
		\text{E}_{T,K} \left[ \,\left| \mu^{(\ro)} w - c(T,K;\alpha) - \mu^{(\text{imp})}(T,K) \right| \,\right],
\end{equation}
where $c(T,K;\alpha)$ is a model for the cost having the parameter set $\alpha$.

\singlespacing
\normalem

\end{document}